\documentclass[10pt,journal,epsfig]{IEEEtran}

\usepackage[dvips]{graphicx}
\usepackage{graphicx}
\usepackage{amssymb}
\usepackage{cite}
\usepackage{subfigure}
\usepackage{amsmath}

\begin{document}

\title{A Fast Iterative Algorithm for Recovery of Sparse Signals from One-Bit Quantized Measurements}

\author{Jun Fang, Yanning Shen, and Hongbin Li,~\IEEEmembership{Senior Member,~IEEE}
\thanks{Jun Fang, and Yanning Shen are with the National Key Laboratory on Communications,
University of Electronic Science and Technology of China, Chengdu
611731, China, Emails: JunFang@uestc.edu.cn,
201121260110@std.uestc.edu.cn}
\thanks{Hongbin Li is
with the Department of Electrical and Computer Engineering,
Stevens Institute of Technology, Hoboken, NJ 07030, USA, E-mail:
Hongbin.Li@stevens.edu}
\thanks{This work was supported in part by the National Science
Foundation of China under Grant 61172114. }}

\maketitle






\begin{abstract}
This paper considers the problem of reconstructing sparse or
compressible signals from one-bit quantized measurements. We study
a new method that uses a log-sum penalty function, also referred
to as the Gaussian entropy, for sparse signal recovery. Also, in
the proposed method, sigmoid functions are introduced to quantify
the consistency between the acquired one-bit quantized data and
the reconstructed measurements. A fast iterative algorithm is
developed by iteratively minimizing a convex surrogate function
that bounds the original objective function, which leads to an
iterative reweighted process that alternates between estimating
the sparse signal and refining the weights of the surrogate
function. Connections between the proposed algorithm and other
existing methods are discussed. Numerical results are provided to
illustrate the effectiveness of the proposed algorithm.
\end{abstract}


\section{Introduction}
Conventional compressed sensing framework recovers a sparse signal
$\boldsymbol{x}\in\mathbb{R}^{n}$ from only a few linear
measurements:
\begin{align}
\boldsymbol{y}=\boldsymbol{A}\boldsymbol{x}
\end{align}
where $\boldsymbol{y}\in\mathbb{R}^{m}$ denotes the acquired
measurements, $\boldsymbol{A}\in\mathbb{R}^{m\times n}$ is the
sampling matrix, and $m\ll n$. Such a problem has been extensively
studied and a variety of algorithms that provide consistent
recovery performance guarantee were proposed, e.g.
\cite{CandesTao05,TroppGilbert07,Wainwright09}. In practice,
however, measurements have to be quantized before being further
processed. Moreover, in distributed systems where data acquisition
is limited by bandwidth and energy constraints, aggressive
quantization strategies which compress real-valued measurements
into one or only a few bits of data are preferred. This has
inspired recent interest in studying compressed sensing based on
quantized measurements. Specifically, in this paper, we are
interested in an extreme case where each measurement is quantized
into one bit of information
\begin{align}
\boldsymbol{b}=\text{sign}(\boldsymbol{y})=\text{sign}(\boldsymbol{A}\boldsymbol{x})
\label{eq1}
\end{align}
where ``$\text{sign}$'' denotes an operator that performs the sign
function element-wise on the vector, the sign function returns $1$
for positive numbers and $-1$ otherwise. Clearly, in this case,
only the sign of the measurement is retained while the information
about the magnitude of the signal is lost. This makes an exact
reconstruction of the sparse signal $\boldsymbol{x}$ impossible.
Nevertheless, if we impose a unit-norm on the sparse signal, it
has been shown \cite{JacquesLaska11,PlanVershynin11} that signals
can be recovered with a bounded error from one bit quantized data.
Besides, in many practical applications such as source
localization, direction-of-arrival estimation, and chemical agent
detection, it is the locations of the nonzero components of the
sparse signal, other than the amplitudes of the signal components,
that have significant physical meanings and are of our ultimate
concern. Recent results \cite{WimalajeewaVarshney12} show that
asymptotic reliable recovery of the support of sparse signals is
possible even with only one-bit quantized data.


The problem of recovering a sparse or compressible signal from
one-bit measurements was firstly introduced by Boufounos and
Baraniuk in their work \cite{BoufounosBaraniuk08}. Following that,
the reconstruction performance from one-bit measurements was more
thoroughly studied
\cite{JacquesLaska11,PlanVershynin11,WimalajeewaVarshney12,LaskaBaraniuk12}
and a variety of one-bit compressed sensing algorithms such as
binary iterative hard thresholding (BIHT) \cite{JacquesLaska11},
matching sign pursuit (MSP) \cite{Boufounos09}, $l_1$
minimization-based linear programming (LP) \cite{PlanVershynin11},
and restricted-step shrinkage (RSS) \cite{LaskaWen11} were
proposed. Although achieving good reconstruction performance,
these algorithms either require the knowledge of the sparsity
level \cite{JacquesLaska11,Boufounos09} or are $l_1$-based methods
that often yield solutions that are not necessarily the sparsest
\cite{PlanVershynin11,LaskaWen11}. In this paper, we study a new
method that uses the Gaussian entropy-based penalty function for
sparse signal recovery. The Gaussian entropy has the potential to
be much more sparsity-encouraging than the $l_1$ norm. By
resorting to a bound optimization approach, we propose an
iterative reweighted algorithm that successively minimizes a
sequence of convex surrogate functions with its weights for the
next iteration computed based on the current estimate. The
proposed algorithm has the advantage that it does not need the
cardinality of the support set, $K$, of the sparse signal.
Moreover, our analysis and simulation results show that the
proposed algorithm outperforms the $l_1$-type methods
\cite{PlanVershynin11,LaskaWen11} in identifying the support set
of the sparse signal.




\section{One-Bit Compressed Sensing}
Since the only information we have about the original signal is
the sign of the measurements, we hope that the reconstructed
signal $\boldsymbol{\hat{x}}$ yields estimated measurements that
are consistent with our knowledge, that is
\begin{align}
\text{sign}(\boldsymbol{a}_i^T\boldsymbol{\hat{x}})=b_i \qquad
\forall i
\end{align}
or in other words
\begin{align}
b_i\boldsymbol{a}_i^T\boldsymbol{\hat{x}}\geq 0 \qquad \forall i
\end{align}
where $\boldsymbol{a}_i$ denotes the transpose of the $i$th row of
the sampling matrix $\boldsymbol{A}$, $b_i$ is the $i$th element
of the sign vector $\boldsymbol{b}$. This consistency can be
enforced by hard constraints \cite{PlanVershynin11,LaskaWen11} or
can be quantified by a well-defined metric which is meant to be
maximized/minimized \cite{Boufounos09,JacquesLaska11,YanYang12}.
In this paper, we introduce the sigmoid function to quantify the
consistency between what we acquired and what we estimated. The
metric is defined as
\begin{align}
\phi(\boldsymbol{x})\triangleq \sum_{i=1}^m
\log(\sigma(b_i\boldsymbol{a}_i^T\boldsymbol{x}))
\end{align}
where
\begin{align}
\sigma(x)\triangleq\frac{1}{1+\exp(-x)} \nonumber
\end{align}
is the sigmoid function. The sigmoid function, with an `S' shape,
approaches one for positive $x$ and zero for negative $x$. Hence
it is a good measure to evaluate the consistency between $b_i$ and
$\boldsymbol{a}_i^T\boldsymbol{x}$. Also, the sigmoid function,
differentiable and log-concave, is more amiable for algorithm
development than the indicator function adopted in
\cite{Boufounos09,JacquesLaska11,YanYang12}. Note that the sigmoid
function, also referred to as the logistic regression model, has
been widely used in statistics and machine learning to represent
the posterior class probability.



Naturally our objective is to find $\boldsymbol{x}$ to maximize
the consistency between the acquired data and the reconstructed
measurements, i.e.
\begin{align}
\max_{\boldsymbol{x}} \qquad \phi(\boldsymbol{x})= \sum_{i=1}^m
\log(\sigma(b_i\boldsymbol{a}_i^T\boldsymbol{x}))
\end{align}
This optimization, however, does not necessarily lead to a sparse
solution. To obtain sparse solutions, a sparsity-encouraging term
needs to be incorporated to encourage sparsity of the signal
coefficients. The most commonly used sparsity-encouraging penalty
function is $l_1$ norm. An attractive property of the $l_1$ norm
is its convexity, which makes the $l_1$-based minimization a
well-behaved numerical problem. Despite its popularity, $l_1$ type
methods suffer from the drawback that the global minimum does not
necessarily coincide with the sparsest solution, particularly when
only a few measurements are available for signal reconstruction.
In this paper, we consider the use of an alternative
sparsity-encouraging penalty function for sparse signal recovery.
This penalty function, also referred to as the Gaussian entropy,
is defined as
\begin{align}
h_G(\boldsymbol{x})=\sum_{i=1}^n \log |x_i|^2 \label{eq2}
\end{align}
where $x_i$ denotes the $i$th component of the vector
$\boldsymbol{x}$. Such a log-sum penalty function was firstly
introduced in \cite{CoifmanWickerhauser92} for basis selection.
This penalty function has the potential to be much more
sparsity-encouraging than the $l_1$ norm. It can be readily
observed that the log-sum (Gaussian entropy) penalty function,
like $l_0$ norm, has infinite slope at $x_i=0,\forall i$, which
implies that a relatively large penalty is placed on small nonzero
coefficients to drive them to zero. The reason why the Gaussian
entropy is more sparsity-encouraging than $l_1$ function will also
be explained from an algorithmic perspective later in our paper.
Using this penalty function, the problem of finding a sparse
solution to maximize the consistency can be formulated as follows
\begin{align}
\hat{\boldsymbol{x}}&=\arg \min L(\boldsymbol{x})\nonumber\\
&=\arg \min_{\boldsymbol{x}}-\sum_{i=1}^m\log(\sigma(b_i
\boldsymbol{x}^T \boldsymbol{a}_i))
+\lambda\sum_{i=1}^n\log|x_i|^2  \label{opt1}
\end{align}
where $\lambda$ is a parameter controlling the trade-off between
the degree of sparsity and the quality of consistency.


\subsection{Proposed Iterative Algorithm}
Due to the concave and unbounded nature of the Gaussian entropy,
the objective function (\ref{opt1}) is non-convex and unbounded
from below. Instead of resorting to the gradient descend method,
we propose an iterative reweighted algorithm which provides fast
convergence and guarantees that every iteration results in a
decreasing objective function value. The algorithm is developed
based on a bound optimization approach \cite{LangeHunter00}. The
idea is to construct a surrogate function
$Q(\boldsymbol{x}|\boldsymbol{\hat{x}}^{(t)})$ such that
\begin{align}
Q(\boldsymbol{x}|\boldsymbol{\hat{x}}^{(t)})-L(\boldsymbol{x})\geq
0
\end{align}
and the minimum is attained when
$\boldsymbol{x}=\boldsymbol{\hat{x}}^{(t)}$, i.e.
$Q(\boldsymbol{\hat{x}}^{(t)}|\boldsymbol{\hat{x}}^{(t)})=L(\boldsymbol{\hat{x}}^{(t)})$.
Optimizing $L(\boldsymbol{x})$ can therefore be replaced by
minimizing the surrogate function
$Q(\boldsymbol{x}|\boldsymbol{\hat{x}}^{(t)})$ iteratively.
Suppose that
\begin{align}
\boldsymbol{\hat{x}}^{(t+1)}=\min_{\boldsymbol{x}}
Q(\boldsymbol{x}|\boldsymbol{\hat{x}}^{(t)}) \nonumber
\end{align}
We can ensure that
\begin{align}
L(\boldsymbol{\hat{x}}^{(t+1)})=&L(\boldsymbol{\hat{x}}^{(t+1)})-Q(\boldsymbol{\hat{x}}^{(t+1)}|\boldsymbol{\hat{x}}^{(t)})
+Q(\boldsymbol{\hat{x}}^{(t+1)}|\boldsymbol{\hat{x}}^{(t)})
\nonumber\\
< &
L(\boldsymbol{\hat{x}}^{(t)})-Q(\boldsymbol{\hat{x}}^{(t)}|\boldsymbol{\hat{x}}^{(t)})
+Q(\boldsymbol{\hat{x}}^{(t+1)}|\boldsymbol{\hat{x}}^{(t)})
\nonumber\\
\leq &
L(\boldsymbol{\hat{x}}^{(t)})-Q(\boldsymbol{\hat{x}}^{(t)}|\boldsymbol{\hat{x}}^{(t)})
+Q(\boldsymbol{\hat{x}}^{(t)}|\boldsymbol{\hat{x}}^{(t)})
\nonumber\\
=& L(\boldsymbol{\hat{x}}^{(t)}) \label{inequality}
\end{align}
where the first inequality follows from the fact that
$Q(\boldsymbol{x}|\boldsymbol{\hat{x}}^{(t)})-L(\boldsymbol{x})$
attains its minimum when
$\boldsymbol{x}=\boldsymbol{\hat{x}}^{(t)}$; the second inequality
comes by noting that
$Q(\boldsymbol{x}|\boldsymbol{\hat{x}}^{(t)})$ is minimized at
$\boldsymbol{x}=\boldsymbol{\hat{x}}^{(t+1)}$. We see that,
through minimizing the surrogate function iteratively, the
objective function $L(\boldsymbol{x})$ is guaranteed to decrease
at each iteration.

We now discuss how to find a surrogate function for the original
problem (\ref{opt1}). Ideally, we hope that the surrogate function
is differentiable and convex so that the minimization of the
surrogate function is a well-behaved numerical problem. Since the
consistency evaluation term is convex, our objective is to find a
convex surrogate function for the Gaussian entropy defined in
(\ref{eq2}). An appropriate choice of such a surrogate function
has a quadratic form and is given by
\begin{align}
f(\boldsymbol{x}|\boldsymbol{\hat{x}}^{(t)})\triangleq\sum_{i=1}^n
\bigg(\frac{x_i^2}{(\hat{x}_i^{(t)})^2}+\log(\hat{x}_i^{(t)})^2 -1
\bigg)
\end{align}
It can be easily verified that
\begin{align}
f(\boldsymbol{x}|\boldsymbol{\hat{x}}^{(t)})-h_G(\boldsymbol{x})\geq
0
\end{align}
with the minimum $0$ attained when
$\boldsymbol{x}=\boldsymbol{\hat{x}}^{(t)}$. Therefore the convex
function $f(\boldsymbol{x}|\boldsymbol{\hat{x}}^{(t)})$ is a
desired surrogate function for the Gaussian entropy
$h_G(\boldsymbol{x})$. As a consequence, the surrogate function
for the objective function $L(\boldsymbol{x})$ is given by
\begin{align}
&Q(\boldsymbol{x}|\boldsymbol{\hat{x}}^{(t)}) \nonumber\\
=& -\sum_{i=1}^m\log(\sigma(b_i \boldsymbol{x}^T
\boldsymbol{a}_i)) +\lambda \sum_{i=1}^n
\bigg(\frac{x_i^2}{(\hat{x}_i^{(t)})^2}+\log(\hat{x}_i^{(t)})^2 -1
\bigg) \nonumber\\
=& -\sum_{i=1}^m\log(\sigma(b_i \boldsymbol{x}^T
\boldsymbol{a}_i))
+\lambda\boldsymbol{x}^T\boldsymbol{D}(\boldsymbol{\hat{x}}^{(t)})\boldsymbol{x}+\text{constant}
\label{surrogate-function}
\end{align}
where
\begin{align}
\boldsymbol{D}(\boldsymbol{\hat{x}}^{(t)})\triangleq\text{diag}\left\{(\hat{x}_1^{(t)})^{-2},\ldots,(\hat{x}_n^{(t)})^{-2}\right\}
\nonumber
\end{align}

Optimizing $L(\boldsymbol{x})$ now reduces to minimizing the
surrogate function $Q(\boldsymbol{x}|\boldsymbol{\hat{x}}^{(t)})$
iteratively. For clarity, the iterative algorithm is briefly
summarized as follows.
\begin{enumerate}
\item Given an initialization $\boldsymbol{\hat{x}}^{(0)}$.
\item At iteration $t>0$, minimize
$Q(\boldsymbol{x}|\boldsymbol{\hat{x}}^{(t)})$, which yields a new
estimate $\boldsymbol{\hat{x}}^{(t+1)}$. Based on this new
estimate, construct a new surrogate function
$Q(\boldsymbol{x}|\boldsymbol{\hat{x}}^{(t+1)})$.
\item Go to Step 2 if
$\|\boldsymbol{\hat{x}}^{(t+1)}-\boldsymbol{\hat{x}}^{(t)}\|_2>\epsilon$,
where $\epsilon$ is a prescribed tolerance value; otherwise stop.
\end{enumerate}


\emph{Remark 1:} The second step involves optimization of the
surrogate function $Q(\boldsymbol{x}|\boldsymbol{\hat{x}}^{(t)})$.
Since the surrogate function is differentiable and convex,
minimization of $Q(\boldsymbol{x}|\boldsymbol{\hat{x}}^{(t)})$ is
a well-behaved numerical problem. Any gradient-based search such
as Newton's method which has a fast convergence rate can be used
and is guaranteed to converge to the global maximum.

\emph{Remark 2:} The above algorithm results in a non-increasing
objective function value of $L(\boldsymbol{x})$. In this manner,
the iterative algorithm eventually converges to a local minimum of
$L(\boldsymbol{x})$. It should be emphasized that the cost
function $L(\boldsymbol{x})$ is non-convex. Therefor it is
important to choose a suitable starting point for the algorithm.
Our simulations suggest that initializing with
$\boldsymbol{\hat{x}}^{(0)}=\boldsymbol{A}^T\boldsymbol{b}$
usually delivers good reconstruction performance. Moreover, we
found from our simulations that, despite starting from different
(randomly generated) initial points, if the number of bits, $m$,
is sufficiently large, the support of the reconstructed sparse
solution is guaranteed to coincide with, or at least be a subset
of, the true support of the sparse signal $\boldsymbol{x}$.

\emph{Remark 3:} The proposed iterative algorithm can be
considered as composed of two alternating steps. First, we
estimate $\boldsymbol{x}$ through minimizing the current surrogate
function $Q(\boldsymbol{x}|\boldsymbol{\hat{x}}^{(t)})$. Second,
based on the estimate of $\boldsymbol{x}$, we update the weights
of the weighted $l_2$ norm penalty of the surrogate function. This
alternating process finally results in sparse solutions. To see
this, note that the weighted $l_2$ norm of $\boldsymbol{x}$ has
their weights specified as $\{(\hat{x}_i^{(t)})^{-2}\}$. The
penalty term has the tendency to decrease these entries in
$\boldsymbol{x}$ whose corresponding weights are large, i.e.,
whose current estimates $\{\hat{x}_i^{(t)}\}$ are already small.
This negative feedback mechanism keeps suppressing these entries
until they reach machine precision and becomes zeros, while
leaving only a few prominent nonzero entries survived to meet the
consistency requirement.

\section{Related Work} \label{sec:relation}
Our developed algorithm has a close connection with the FOcal
Underdetermined System Solver (FOCUSS) algorithm
\cite{GorodnitskyRao97} since the latter algorithm also uses an
iterative reweighted approach to find sparse solutions to
underdetermined systems
$\boldsymbol{y}=\boldsymbol{A}\boldsymbol{x}$. Specifically, at
each iteration, FOCUSS solves a reweighted $l_2$ minimization with
weights $w_i^{(t+1)}=1/|x_i^{(t)}|^p$, where $p\in[1\phantom{0}
2]$. It was also shown that when $p=2$, FOCUSS is equivalent to a
modified Newton's method minimizing the Gaussian entropy function.
As compared with FOCUSS, our paper considers a more general
framework in which the data model allows measurement errors/noise
and is not confined to be linear, and a general connection between
the sparsity-promoting penalty function and the iterative
reweighted process is established through the use of the surrogate
function. In \cite{CandesWakin08}, a similar iterative reweighted
algorithm was proposed to enhance sparsity of recovered signals.
The algorithm consists of solving a sequence of weighted
$l_1$-minimization problems with its weights for the next
iteration computed based on the current estimate. Such an
algorithm, interestingly, is also shown closely related to the
log-sum penalty function, i.e. the Gaussian entropy function.

We now explore the relationship between our proposed algorithm and
the $l_1$-minimization type methods. Following (\ref{opt1}), we
can formulate a $l_1$ norm-based sparsity-promoting optimization
\begin{align}
\hat{\boldsymbol{x}}&=\arg \min_{\boldsymbol{x}} \tilde{L}(\boldsymbol{x})\nonumber\\
&=\arg \min_{\boldsymbol{x}}-\sum_{i=1}^m\log(\sigma(b_i
\boldsymbol{x}^T \boldsymbol{a}_i)) +\lambda\sum_{i=1}^n|x_i|
\label{opt2}
\end{align}
Similarly, a surrogate function can be constructed to bound the
above optimization, from which an iterative algorithm can be
developed to solve (\ref{opt2}). Note that the $l_1$ function
$|x|$ can also be upper bounded by a quadratic function. It can be
readily verified that the surrogate function for $l_1$ norm is
given by
\begin{align}
\tilde{f}(\boldsymbol{x}|\boldsymbol{\hat{x}}^{(t)})\triangleq\frac{1}{2}\sum_{i=1}^n
\bigg(\frac{x_i^2}{|\hat{x}_i^{(t)}|}+|\hat{x}_i^{(t)}|\bigg)
\end{align}
Therefore the solution to (\ref{opt2}) can be found by iteratively
minimizing the following surrogate function
\begin{align}
&\tilde{Q}(\boldsymbol{x}|\boldsymbol{\hat{x}}^{(t)}) \nonumber\\
=& -\sum_{i=1}^m\log(\sigma(b_i \boldsymbol{x}^T
\boldsymbol{a}_i)) +\frac{\lambda}{2} \sum_{i=1}^n
\bigg(\frac{x_i^2}{|\hat{x}_i^{(t)}|}+|\hat{x}_i^{(t)}|
\bigg) \nonumber\\
=& -\sum_{i=1}^m\log(\sigma(b_i \boldsymbol{x}^T
\boldsymbol{a}_i))
+\frac{\lambda}{2}\boldsymbol{x}^T\boldsymbol{\tilde{D}}(\boldsymbol{\hat{x}}^{(t)})\boldsymbol{x}+\text{constant}
\label{surrogate-function2}
\end{align}
where
\begin{align}
\boldsymbol{\tilde{D}}(\boldsymbol{\hat{x}}^{(t)})\triangleq\text{diag}\bigg\{\frac{1}{|\hat{x}_1^{(t)}|},\ldots,\frac{1}{|\hat{x}_n^{(t)}|}\bigg\}
\nonumber
\end{align}
The surrogate function
$\tilde{Q}(\boldsymbol{x}|\boldsymbol{\hat{x}}^{(t)})$ has a
similar form as (\ref{surrogate-function}), except that the
surrogate function for the Gaussian entropy updates its weights
using $(\hat{x}_i^{(t)})^{-2}$, while the surrogate function for
the $l_1$ norm updates its weights with
$(|\hat{x}_i^{(t)}|)^{-1}$. This seemingly slight difference,
however, results in very different convergence behavior. The
latter update equation guarantees converging to a unique global
minimum. On the other hand, the update method used in our
algorithm renders a more intense effect in de-emphasizing the
entries in $\boldsymbol{x}$. Therefore the proposed algorithm is
more sparsity-encouraging than the $l_1$ type methods, and
generally yields a more sparse solution.







\section{Numerical Results}
We now carry out experiments to illustrate the performance of our
proposed one-bit compressed sensing algorithm. In our simulations,
the $K$-sparse signal is randomly generated with the support set
of the sparse signal randomly chosen according to a uniform
distribution. The signals on the support set are independent and
identically distributed (i.i.d.) Gaussian random variables with
zero mean and unit variance. The measurement matrix
$\boldsymbol{A}\in\mathbb{R}^{m\times n}$ is randomly generated
with each entry independently drawn from Gaussian distribution
with zero mean and unit variance, and then each column of
$\boldsymbol{A}$ is normalized to unity for algorithm stability.
We compare our proposed algorithm with the other two algorithms,
namely, the $l_1$ minimization-based linear programming (LP)
algorithm \cite{PlanVershynin11} (referred to as ``one-bit LP''),
and the iterative algorithm proposed in Section \ref{sec:relation}
to solve the optimization (\ref{opt2}) (referred to as
``L1-optimization (\ref{opt2})''). Both the latter two algorithms
are $l_1$ type methods, with the consistency evaluated by
different criteria.

\begin{figure}[!t]
 \centering
\begin{tabular}{cc}
\hspace*{-3ex}
\includegraphics[width=4.8cm]{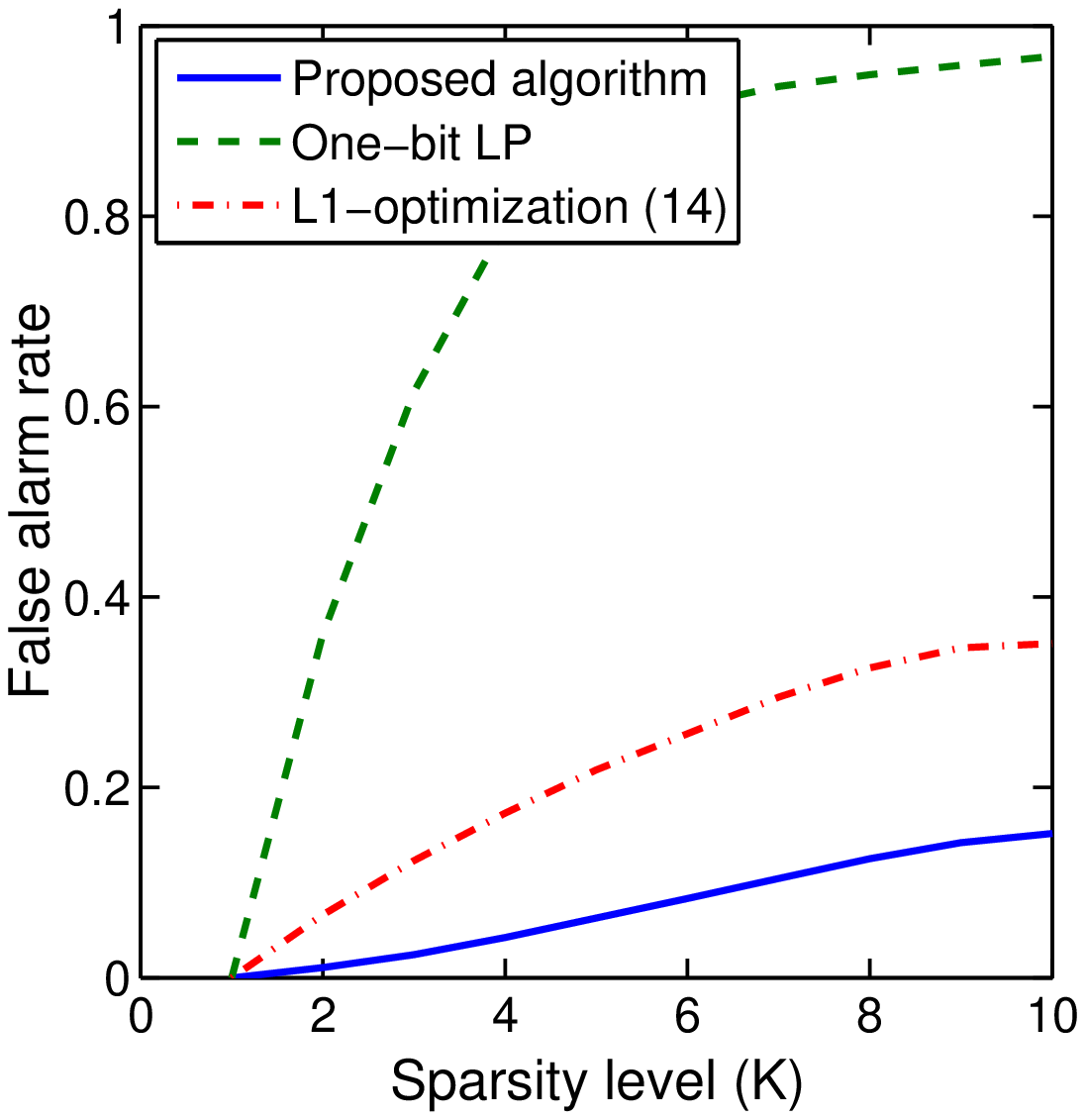}&
\hspace*{-5ex}
\includegraphics[width=4.8cm]{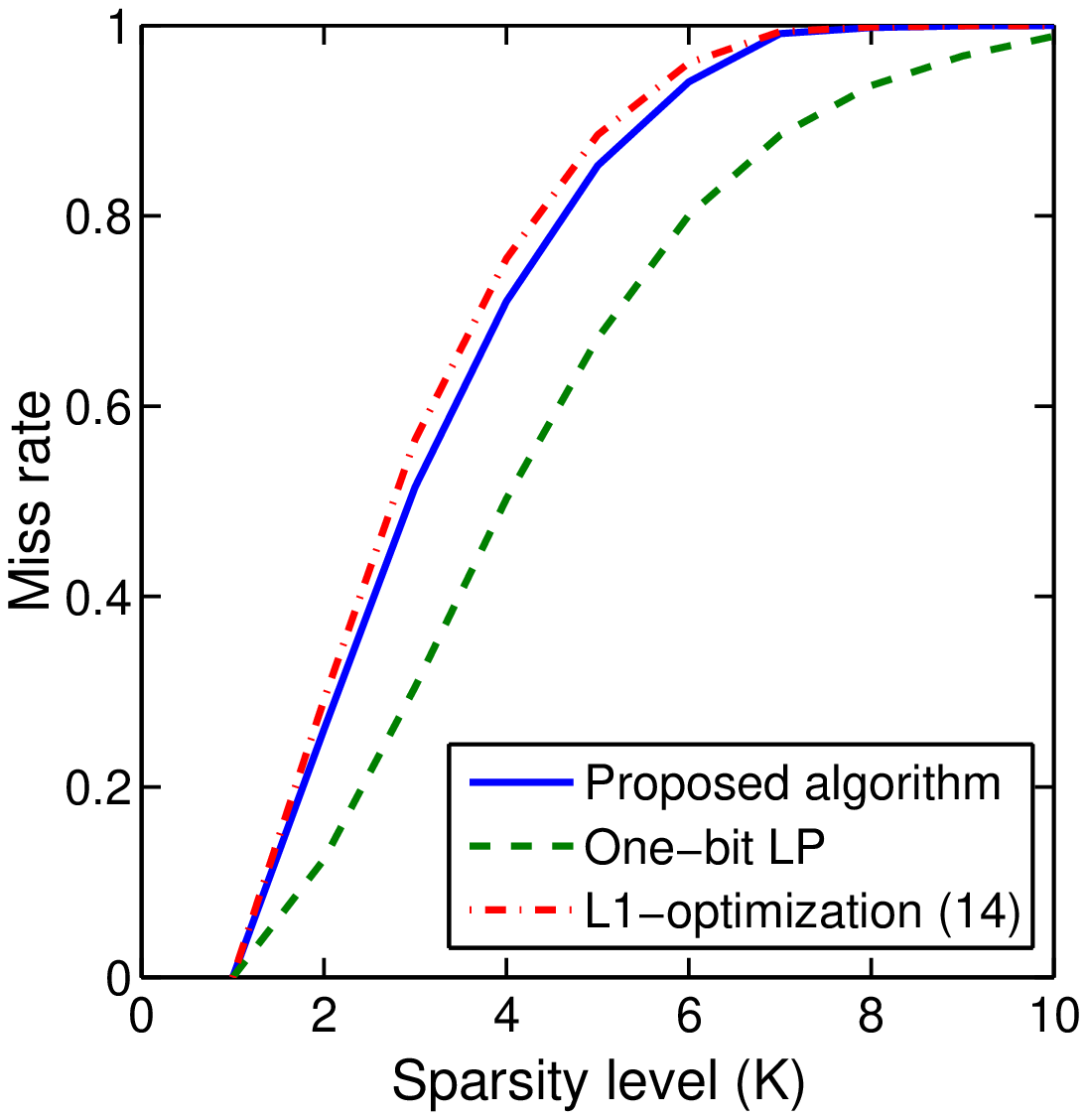}
\\
(a)& (b)
\end{tabular}
  \caption{(a). False alarm rates of respective algorithms; (b). Miss rates of respective algorithms.}
   \label{fig1}
\end{figure}

We investigate the support recovery performance of respective
algorithms. Support recovery accuracy are measured by the false
alarm (misidentified) rate and the miss rate. A false alarm event
represents the case where coefficients that are zero in the
original signal are misidentified as nonzero after reconstruction,
while a miss event stands for the case where the nonzero
coefficients are missed and determined to be zero. Fig. \ref{fig1}
depicts the false alarm and miss rates of respective algorithms as
a function of the sparsity level $K$, where we set $m=100$,
$n=50$, and $\lambda=1/2$ in our simulations. Results are averaged
over $10^4$ independent runs. We see that the proposed algorithm
provides more accurate identification of the true support set: it
has a similar (or slightly higher) miss rate as (than) that of the
$l_1$-based methods, while meanwhile achieves a considerably lower
false alarm rate than both $l_1$ type methods. Our result also
indicates that the proposed algorithm yields solutions more sparse
than $l_1$ type methods, which corroborates our theoretical
analysis.






\section{Conclusions}
We proposed an iterative reweighted algorithm for sparse signal
reconstruction from one-bit quantized measurements. The proposed
algorithm consists of solving a sequence of minimization problems
whose weights are updated based on the current estimate. Analyses
and simulation results show that the proposed algorithm that uses
the log-sum penalty function is more sparsity-encouraging than
$l_1$-based methods, and outperforms $l_1$ type methods in
recovering the true support of the sparse signal.

\useRomanappendicesfalse
\appendices

\bibliography{MyBibTex}
\bibliographystyle{IEEEtran}

\end{document}